\documentclass[aps,prb,floatfix,twocolumn]{revtex4}
\usepackage[pdftex]{graphicx}
\usepackage{amsmath,amssymb,subfigure,color,verbatim}
\usepackage{yllmath}
\providecommand{\mysection}[1]{ \textit{#1}: }

\begin{document}
\title{Accurate Calculation of Off-Diagonal Green Functions on Anisotropic Hypercubic Lattices}
\author{Yen Lee Loh}
\affiliation{Department of Physics and Astrophysics, University of North Dakota, Grand Forks, ND  58202}
\date{2012-10-21} 
\begin{abstract}
We present a method for accurate evaluation of the Green function $G(\omega,r_1,\dotsc,r_d)$ at any real frequency $\omega$ and any lattice vector $(r_1,\dotsc,r_d)$ for a $d$-dimensional hypercubic lattice that may have anisotropic couplings $(\Omega_1,\dotsc,\Omega_d)$. 
In this method, we start with an integral representation of $G$, split the oscillatory integrand into combinations of Hankel functions, and deform the integration paths into the complex plane to obtain rapidly convergent integrals.
We also discuss an alternative approach using the Levin collocation method.
We report values of the Green function at selected frequencies on the branch cut and selected lattice vectors.
\end{abstract}
\maketitle

Lattice Green functions arise in numerous problems in combinatorics, statistical mechanics, and condensed matter physics.
In the language of condensed matter physics, the Green function $G(\rrr,\omega)$ is the propagator for a particle of energy (or frequency) $\omega$ to move a distance $\rrr$ in a tight-binding model on a lattice with a nearest-neighbor hopping.  
Accurate values for lattice Green functions are useful as input for further calculations in many-body theory, such as simulations of Hubbard models\cite{bloch2008review} and non-perturbative renormalization group studies.\cite{caillol2012arxiv}

Closed-form expressions exist for Green functions on many lattices in one, two, and three dimensions, for Bethe lattices, and for certain fractal lattices\cite{schwalm1988,schwalm1992}; see Ref.~\onlinecite{guttmann2010} and references therein.
In particular, Joyce found closed forms for the Green function of the cubic lattice at the origin,\cite{joyce1972,joyce1994} at a general lattice point,\cite{joyce2002} and of an anisotropic cubic lattice.\cite{delves2001}  However, closed forms for hypercubic lattice Green functions in $d\geq 4$ dimensions are not available in terms of named special functions.\cite{joyce2001,joyce2003} 

We recently developed a method\cite{loh2011hlgf} for obtaining accurate numerical values for the on-site Green function $G_\0 (\omega)$ on a $d$-dimensional hypercubic lattice.
In this work we generalize this method to off-diagonal Green functions $G_\rrr (\omega)$ on anisotropic lattices; the generalization is not as trivial as one might expect.  We also discuss an alternative approach using the Levin collocation method to integrate oscillatory functions, and we make an interesting observation about the computational complexity of both methods.  Finally, we report values of the Green function as test cases against which to benchmark implementations or other methods.

Consider a $d$-dimensional hypercubic lattice with anisotropic nearest-neighbour hopping amplitudes $\half \Omega_k$ ($k=1,2,\dotsc,d$) in each direction.  The Green function $G$ has closed forms in various representations involving time $t$, frequency $\omega$, position $\rrr$, and wavevector $\qqq=(q_1, \dotsc, q_d)$:
	\begin{align}
	G(\qqq, \omega) 
	&=\frac{1}{\omega - \Omega_1 \cos q_1 - \dotso - \Omega_d \cos q_d} 
		\label{Gomegak}
		\\
	G(\qqq, t) 
	&=-i \Theta(t) \exp  {it (\Omega_1 \cos q_1 - \dotso - \Omega_d \cos q_d)}
		\\
	G_\rrr (t)
	&=i^{r_1+\dotso+r_d-1} \Theta(t) J_{r_1} (\Omega_1 t) \dotso J_{r_d} (\Omega_d t).
	\end{align}
The position-frequency Green function $G_\rrr (\omega)$ can be written as a $d$-dimensional integral over the Brillouin zone,
	\begin{align}
	G_\rrr (\omega) 
	&=\frac{1}{(2\pi)^d}
		\int_0^{2\pi} dq_1 \dots \int_0^{2\pi} dq_d
				\nonumber\\&~~~~~~{}
		\frac{\exp i(q_1 r_1 + \dotso + q_d r_d)}{\omega - \Omega_1 \cos q_1 - \dotso - \Omega_d \cos q_d + i0^+}
		\label{BZIntegral}
	\end{align}
where we have included an infinitesimal shift in the denominator as is conventional in condensed matter theory.  This form makes it clear that $G_\rrr (\omega)$  has a branch cut along the real axis representing a continuous spectrum (band) of particle excitations.  However, for numerical evaluation of $G_\rrr (\omega)$ it is better to start from the one-dimensional integral representation\cite{maradudin1960,oitmaa1971,loh2011hlgf}
	\begin{align}
	G_\rrr (\omega)
	&=i^\alpha
		\int_0^\infty dt~ e^{i\omega t} J_{r_1} (\Omega_1 t) \dotso J_{r_d} (\Omega_d t)
	\label{BesselFormula}
	\end{align}
where $	\alpha=r_1+\dotso+r_d-1$ and $\{r_k\}$ are integers.

	\begin{figure*}[!htb]
		\includegraphics[width=0.99\textwidth]{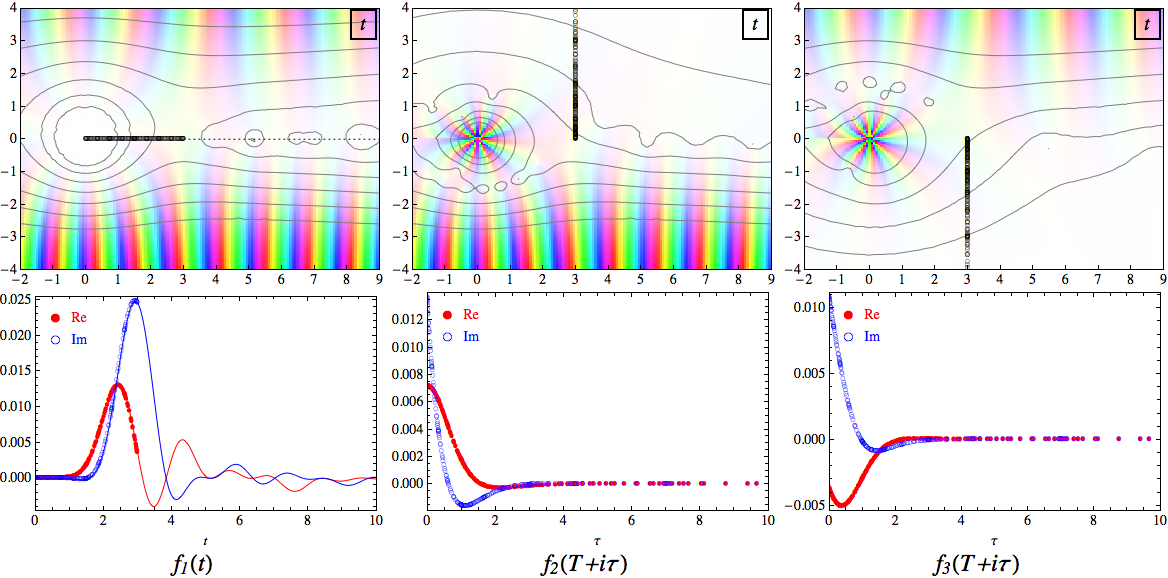}	
	\caption{
	\label{Argand}
	(Color online)
  Upper panels show integrands for the evaluation of $G_\rrr(\omega)$ 
  	using the complex-plane integration technique.
	Parameters were
		$d=4$, $\Omega_k=\{1,1,1,1\}$, $\omega=1$, $r_k=\{1,2,2,3\}$.
	The split point was $T=3$.
	Hue indicates phase and saturation indicates modulus.
	Curves are equal-modulus contours $\left|f\right|=10^{-4}, 10^{-3}, \dotsc, 10^2$.
	Black circles indicate quadrature points along integration path.
	Lower panels show behavior of each integrand along its integration path;
		symbols indicate evaluation points.
	It took
	165 evaluations of $f_1$, 
	253 evaluations of $f_2$, and
	253 evaluations of $f_3$ to reach 12-digit accuracy 
	using an adaptive Gauss-Kronrod quadrature rule.
	For comparison, the original integral formula Eq.~\eqref{BesselFormula}
	was unable to achieve 6-digit accuracy even after 50000 function evaluations.
	}
	\end{figure*}

\section{Complex-plane method} 
The integrand of Eq.~\eqref{BesselFormula} is analytic at $t=0$, but as $t\rightarrow \infty$ it exhibits irregular oscillations with a slowly decaying envelope ($t^{-d/2}$), which causes problems with traditional quadrature schemes.  For example, in $d=4$ the integrand decays as $t^{-2}$, so it becomes negligible (i.e., below machine precision) only beyond $t \sim 10^8$.  Therefore it is useful to adopt the complex-plane technique of Ref.~\onlinecite{loh2011hlgf}.

Split the integral at a point $t=T$ on the real axis by writing
	\begin{align}
	G_\rrr(\omega) &= g_1 + g_2
	\label{gsplit}
		,\\
	g_1
	&=i^\alpha
	 \int_0^T dt~ e^{i\omega t} J_{r_1} (\Omega_1 t) \dotso J_{r_d} (\Omega_d t)
	\label{g1}
		,\\
	g_2
	&=i^\alpha
		\int_T^\infty dt~ e^{i\omega t} J_{r_1} (\Omega_1 t) \dotso J_{r_d} (\Omega_d t)
	\label{g2}
		.
	\end{align}
In Eq.~\eqref{g2}, apply the identity
	\begin{align}
	J_r(u) &= \half \left[  H^+_r(u) + H^-_r(u)  \right]
	\end{align}
where $H^+_r(u) \equiv H^{(1)}_r(u)$ and $H^-_r(u) \equiv H^{(2)}_r(u) \equiv [H^{(1)}_r(u^*)]^*$ are Hankel functions of the first and second kind respectively.  Performing a binomial expansion gives a sum of $2^d$ combinations of Hankel functions, so that
	\begin{align}
	g_2
	&=\frac{i^\alpha}{2^d}
		\sum_{ \{\sigma\} }
		 \int_T^\infty dt~ e^{i\omega t} 
			\prod_k  H^{\sigma_k}_{r_k} (\Omega_k t)
		,
	\end{align}
where the sum over $\{\sigma\}$ runs over all $2^d$ configurations of the Ising variables $\sigma_k = \pm 1$ and the sum over $k$ runs from 1 to $d$.  For $\abs{t} \rightarrow \infty$ the Hankel functions behave as $H^\sigma_r(u) \sim e^{i \sigma u}$.  Thus, for a given $\{\sigma\}$, the integrand goes as $e^{i\omega t} e^{i\sigma_1 \Omega_1 t} \dotso  e^{i\sigma_d \Omega_d t}$, and Jordan's lemma allows us to rotate the integration path into the upper or lower half-plane (UHP/LHP) depending on the sign of $\Lambda[\sigma] = \omega + \sigma_1\Omega_1 + \dotso + \sigma_d\Omega_d$.
Thus
	\begin{align}
	g_2
	&=\frac{i^\alpha}{2^d}
		\sum_{ \{\sigma\} }
		\begin{cases}
			\displaystyle
			\int_T^{T+i\infty} dt~ e^{i\omega t} 	\prod_k  H^{\sigma_k}_{r_k} (\Omega_k t)
		 			& \Lambda \geq 0 \\
			\displaystyle
			\int_T^{T-i\infty} dt~ e^{i\omega t} 	\prod_k  H^{\sigma_k}_{r_k} (\Omega_k t)
		 			& \Lambda < 0 . 
		\end{cases}
	\label{g2northsouth}
	\end{align}
Together, Eqs.~\eqref{gsplit}, \eqref{g1}, and \eqref{g2northsouth} give a prescription for computing $G_\rrr(\omega)$.  

To appreciate why this procedure is necessary and effective, it is instructive to write the Green function in the form
	\begin{align}
	G_\rrr(\omega)
	&=\int_0^T \!\! dt~ f_1(t)
	+ \int_T^{T+i\infty} \!\! dt~ f_2(t)
	+ \int_T^{T-i\infty} \!\! dt~ f_3(t)
	\label{threeIntegrals}
	\end{align}
where 
	\begin{align}
	f_1(t) &= i^\alpha e^{i\omega t} J_{r_1} (\Omega_1 t) \dotso J_{r_d} (\Omega_d t)
		,\\
	f_2(t) & = \frac{i^\alpha e^{i\omega t}}{2^d} 
			 \sum_{\{\sigma\}}^{\Lambda[\sigma] \geq 0}
			\prod_k  H^{\sigma_k}_{r_k} (\Omega_k t)
		,\\
	f_3(t) & = \frac{i^\alpha e^{i\omega t}}{2^d} 
			 \sum_{\{\sigma\}}^{\Lambda[\sigma] < 0}
			\prod_k  H^{\sigma_k}_{r_k} (\Omega_k t)
		,
	\label{f1f2f3}
	\end{align}
and to visualize these three integrations as in Fig.~\ref{Argand}.  The original integrand $f_1(t)$ oscillates irregularly, whereas $f_2(\tau)$ and $f_3(\tau)$ decay quickly with only a couple of oscillations.

In order to use standard quadrature routines, it is convenient to parametrize the path of integration by a real variable $\tau$.  Returning to Eq.~\eqref{g2northsouth} and substituting $t = T \pm i\tau$ leads to
	\begin{align}
	g_2
	&=\int_0^\infty d\tau~ f_4(\tau)
		,\nonumber\\
	f_4(\tau) &=\frac{i^{\alpha+1} e^{i \omega T}}{2^d} 
		\sum_{ \{\sigma\} }
		\begin{cases}
			\displaystyle
			e^{-\omega \tau} 	\prod_k  \calH^{\sigma_k}_{k}
		 			& \Lambda \geq 0 \\
			\displaystyle
			-e^{+\omega \tau} 	\prod_k  \left( \calH^{-\sigma_k}_{k} \right)^*
		 			& \Lambda < 0
		\end{cases}
	\label{g2tauintegrals}
	\end{align}
where the $2d$ numbers $\{\calH\}$ are 
	\begin{align}
	\calH^{\pm}_{k} &= H^{\pm}_{r_k} \big(\Omega_k (T + i\tau)\big)
	,\\
	(\calH^{\mp}_{k})^* &= H^{\pm}_{r_k} \big(\Omega_k (T - i\tau)\big)
	.
	\end{align}
A na{\"i}ve implementation of Eq.~\eqref{g2tauintegrals} requires evaluating a sum of $2^d$ products each containing $d$ Hankel functions.  For an efficient implementation, one should precompute the two real exponentials and the $2d$ complex Hankel functions, as described in the pseudocode below:
\\

{\tt
Tabulate $X_\pm=e^{\pm \omega\tau}$

Tabulate $\calH^{\pm}_{k} = H^{\pm}_{r_k} \big(\Omega_k (T + i\tau)\big)$
for $k=1,\dotsc,d$

Set $s=0$

For each of the $2^d$ Ising configurations {$\sigma$}

~~~~If $\omega+\sigma_1\Omega_1+\dotso+\sigma_d\Omega_d \geq 0$

~~~~~~~~Set $s = s + X_- 	\prod_k  \calH^{\sigma_k}_{k}$

~~~~Else

~~~~~~~~Set $s = s - X_+ 	\prod_k  (\calH^{-\sigma_k}_{k} )^*$

~~~~End If

End For
}
\\

In order to develop a robust implementation, one should consider the following four situations:

\paragraph{Frequency outside band (Green function not on branch cut):}  
If $\omega \geq \Omega_1 + \dotso + \Omega_d$, then the integrand of Eq.~\eqref{BesselFormula} is dominated by the $e^{i\omega t}$ term, which decays into the UHP.  In this case, one can simply rotate the path of integration to lie along the positive imaginary axis and substitute $t=i\tau$ to obtain
	\begin{align}
	G_\rrr(\omega) &= (-1)^{r_1 + \dotsc + r_d} 
		\int_0^\infty d\tau~  e^{-\omega \tau}  
		I_{r_1} (\Omega_1 \tau) \dotso
		I_{r_d} (\Omega_d \tau) .
	\label{BesselIFormula}
	\end{align}
This method was put forth in Ref.~\onlinecite{maradudin1960}; see also Ref.~\onlinecite{delves2001}.

\paragraph{Generic frequency within band:}  
In general, the integrand $f_4(\tau)$ decays exponentially as $\tau\rightarrow \infty$.  This case is well handled by standard numerical integration routines.

\paragraph{Frequency at van Hove singularity:}  
The Green function $G_\rrr(\omega)$ has singularities at frequencies $\pm \Omega_1 \pm \Omega_2 \pm \dotso \pm \Omega_d$.  For a generic anisotropic lattice there are $2^d$ singularities, whereas for an isotropic lattice with $\Omega_k=1$ there are only $(d+1)$ distinct singularities at $\{-d,-d+2,\dotsc,d\}$.  Whenever $\omega$ lies at one of these van Hove singularities ($\omega_c$), $f_4(\tau)$ decays only as a power law, $\tau^{-d/2}$.  This poses no fundamental problem; however, canned integration routines that do automatic singularity handling may attempt to evaluate the Hankel functions at very large $\omega$, causing overflow or underflow.  Therefore, it is advantageous to ``fold'' the integration domain manually as follows.  Split the integral in Eq.~\eqref{g2tauintegrals} at a point $\tau=\eta$ and apply the transformation $\tau = u^{2/(2-d)}$ to the second term to obtain
	\begin{align}
	g_2
	&=\int_0^\eta d\tau~ f_4(\tau)
	+ \tfrac{2}{d-2}
		\int_0^{\eta^{1-d/2}} du~   u^{d/(2-d)} f_4(  u^{2/(2-d)}  )
		.
	\label{g2PowerLawTrick}
	\end{align}
The Jacobian cancels the asymptotic $\tau^{-d/2}$ behavior, so the transformed integrand tends to a constant as $u\rightarrow 0$, and can be integrated easily using conventional quadrature routines.

\paragraph{Frequency near van Hove singularity:}  
If $\omega \approx \omega_c$, the integrand $f_4(\tau)$ crosses over from power law decay to exponential decay at a large value of the argument, $\tau \sim 1/\left| \omega - \omega_c \right|$.  Any adaptive quadrature routine will be forced to sample $f_4(\tau)$ in the vicinity of this crossover, causing overflow and underflow.  In this situation it may be necessary to rewrite the integrand in terms of exponentially scaled Hankel functions $e^{\mp iz} H^\pm_r(z)$, which may be computed directly from appropriate recursions or from asymptotic expansions such as
	\begin{align}
	e^{-iz} H^+_r (z)
	&=\sqrt{\tfrac{2}{\pi z}}  \exp i(-\tfrac{\pi}{4} - \tfrac{r\pi}{2}) + \dotso
	.
	\label{ExponentiallyScaledHankel}
	\end{align}

There is some flexibility in choosing the split point $T$.  With $T=\infty$, one recovers the original formula Eq.~\eqref{BesselFormula}.  Taking $T=0$ corresponds to integrating $f_2(t)$ from $0$ to $+i\infty$ and $f_3(t)$ from $0$ to $-i\infty$ respectively.  This is an attractive choice because Hankel functions of imaginary argument can be computed cheaply by decomposing them into Bessel $K$ and Bessel $I$ functions.\cite{loh2011hlgf}  However, although the original integrand $f_1(t)$ was analytic at $t=0$, the split integrands $f_2(t)$ and $f_3(t)$ are singular at $t=0$.  In the case of the on-site Green function ($\rrr=\0$),  $f_2(t)$ and $f_3(t)$ have logarithmic branch points at $t=0$, which are weak, integrable singularities, so it is permissible to rotate the integration paths as was done in Ref.~\onlinecite{loh2011hlgf}.  However, for $\rrr\neq \0$, $f_2(t)$ and $f_3(t)$ both have a pole at $t=0$ (which may be a high-order pole), so that it is illegal to rotate the path about $t=0$.  Then it is necessary to use a split point $T>0$ sufficiently far away from the pole at $t=0$.

The optimal value of $T$ depends on the implementation of Bessel and Hankel functions and the quadrature routines available in one's software package.  In the current version of \emph{Mathematica} at the time of writing (8.0.1.0), the special function routines are slow (\texttt{HankelH1[0,2.+3.I]} takes almost 0.7 milliseconds) and inaccurate (the error in \texttt{BesselJ[0,11.]} is 1000 times larger than machine precision), so we have not attempted detailed fine-tuning.

There is also some flexibility in the choice of integration paths.  Instead of integrating $f_2(t)$ along the straight line $T \longrightarrow T+i\infty$, it may be desirable to use the path $T \longrightarrow iT \longrightarrow i\infty$, because the integrand is cheaper to evaluate when $t$ is pure imaginary.

For general $\rrr$, the binomial expansion leads to $2^d$ terms.  In certain cases these are not all distinct.  For example, for $\rrr=\0$ on an isotropic lattice, the binomial expansion of $J_0(t)^d$ only gives rise to $(d+1)$ distinct terms \cite{loh2011hlgf}.

\section{Levin collocation method} 
As remarked earlier, the integrand of Eq.~\eqref{BesselFormula} has irregular oscillations that make it difficult to integrate using traditional quadrature schemes.
However, integrands of this form are amenable to relatively modern integration techniques such as the Levin collocation method \cite{levin1982,levin1996}.  Mathematica 8.0.1.0 is able to perform an automatic symbolic analysis of the integrand to determine a suitable set of basis functions and carry out the Levin method.  We summarize the method below, assuming the summation convention for brevity.
 
The Levin method deals with an integral of the form $I = \int_a^b dt~ f_i w_i$ (summed over $i=1,2,3,\dotsc,n$) where $f_i(t)$  are slowly varying envelope functions and $w_i(t)$ are oscillatory basis functions.  The $n$ basis functions satisfy a matrix ordinary differential equation (ODE), $w_i' = A_{ij} w_j$, where $A_{ij}(t)$ vary slowly.  The fundamental theorem of calculus gives the integral in terms of the antiderivative of the integrand:
	\begin{align}
	I &= F_i(b) w_i(b) - F_i(a) w_i(a)	
	\label{IFromFi}
	\end{align}
where $(F_i w_i)' = f_i w_i$.  Simple manipulations show that the antiderivative envelope functions $F_i(t)$ satisfy the matrix ODE
	\begin{align}
	F_i' + A_{ji} F_j = f_i,
	\label{FiODE}
	\end{align}
in which the ``kernel'' $A_{ji}(t)$ and ``forcing'' $f_i(t)$ vary slowly.

The general solution for $F_i(t)$ contains an admixture of up to $n$ oscillatory ``modes''.  However, Levin observed that there is \emph{one} solution that is slowly varying, and that this ``particular solution'' can be well approximated by solving the ODE using a collocation method as follows.  Choose a set of collocation points $\{t_l\}$ for $l=0,1,2,\dotsc,m-1$.  It is convenient to use the extrema of the Chebyshev polynomial of degree $m$ (suitably scaled): 
$
	t_l = \frac{1}{2} [b+a - (b-a) \cos \frac{l\pi}{m-1}]
	.
$
Choose an ansatz for $F_i(t)$.  It is convenient to use a linear combination of Chebyshev polynomials of degree less than $m$, weighted by coefficients $c_{ik}$ ($k=0,1,2,\dotsc,m-1$),
	\begin{align}
	F_i(t) &\approx c_{ik} u_k(t)
		\text{~where~}
	u_k(t) = T_k \!\big( \tfrac{2t-b-a}{b-a} \big).
	\label{FiAnsatz}
	\end{align}
The derivatives are series in Chebyshev polynomials of the second kind, $U_k$:
	\begin{align}
	F_i'(t) &\approx c_{ik} u_k'(t)
		\text{~where~}
	u_k'(t) = \tfrac{2k}{b-a} U_{k-1} \!\big( \tfrac{2t-b-a}{b-a} \big).
	\label{FiPrimeAnsatz}
	\end{align}
Let us demand that the ansatz satisfy the ODE exactly at the collocation points.  Putting Eqs.~\eqref{FiAnsatz} and \eqref{FiPrimeAnsatz} into Eq.~\eqref{FiODE} leads to
	\begin{align}
	\left[ 
		u_k'(t_l) \delta_{ij} + 
		A_{ji} (t_l) 
		u_k(t_l)
	\right] c_{jk}   &= f_i (t_l).	
	\label{LargeLinearSystem}
	\end{align}
This is a system of $mn$ linear equations, which can be solved to give the $mn$ coefficients $c_{jk}$ in $O(m^3n^3)$ time.  From the numerical values of the $c_{jk}$ one can then use Eq.~\eqref{FiAnsatz} in Eq.~\eqref{IFromFi} to estimate the integral $I$.

	\begin{figure*}[!htb]
		\includegraphics[width=0.99\textwidth]{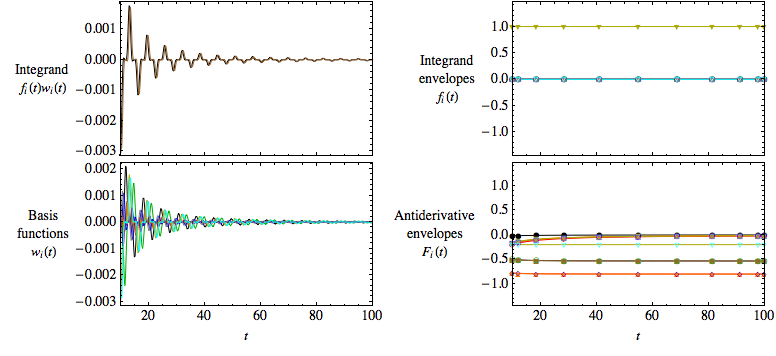}	
	\caption{
	\label{levin}
	(Color online)
	Graphical examination of the Levin method as applied to the integral
	$I=\int_{10}^{100} dt~ e^{1.5i t} J_0(t)^4$.
	The oscillatory integrand
	is written as a combination of $n=5$ basis functions $w_i(t)$
	modulated by integrand envelope functions $f_i(t)$
	(both real and imaginary parts are plotted).
	The antiderivative (\emph{indefinite} integral) is represented 
	in the same basis by antiderivative envelope functions $F_i(t)$.
	The $F_i(t)$ are approximated by Chebyshev polynomials
	and the Chebyshev coefficients $c_{jk}$ are determined by demanding that
	$F_i(t)$ satisfies a matrix ODE at $m=11$ collocation points $t_l$.
	The values of $f_i(t_l)$ and $F_i(t_l)$ are illustrated as circles
	(although the latter are not needed for the calculation of $I$).
	The change in the antiderivative, $F_i(100) w_i(100) - F_i(10) w_i(10)$,
	gives a good approximation to $I$.  (We are using the summation convention.)
	}
	\end{figure*}

Let us now examine the application of the Levin method to the computation of the Green function $G_{r_1r_2\dots r_d}(\omega)$ via Eq.~\eqref{BesselFormula}.  In general, this requires $2^d$ basis functions, $w_i(t)$, which are all possible products of Bessels and Bessel derivatives,
	\begin{align}
	&J_{r_1} (\Omega_1 t) 	J_{r_2} (\Omega_2 t)  \dotso  	J_{r_d} (\Omega_d t),\nonumber\\
	&J'_{r_1} (\Omega_1 t) 	J_{r_2} (\Omega_2 t)  \dotso  	J_{r_d} (\Omega_d t),\nonumber\\
	&J_{r_1} (\Omega_1 t) 	J'_{r_2} (\Omega_2 t)  \dotso  	J_{r_d} (\Omega_d t),\nonumber\\
	&\dots \nonumber\\
	&J'_{r_1} (\Omega_1 t) 	J'_{r_2} (\Omega_2 t)  \dotso  	J'_{r_d} (\Omega_d t).
	\end{align}
For clarity of exposition we illustrate the procedure for the on-site Green function on an isotropic lattice in $d=4$, $G_\0(\omega)=-i\int_0^\infty e^{i\omega t} J_0(t)^4$, which requires only $n=d+1$ basis functions.  The basis, forcings, and differential matrix are
	\begin{align}
	\{w_i\} &= \pmat{  
		 e^{i\omega t} J_0' (t)^4 \\
		 e^{i\omega t} J_0' (t)^3 J_0(t) \\
		 e^{i\omega t} J_0' (t)^2 J_0(t)^2 \\
		 e^{i\omega t} J_0' (t) J_0(t)^3 \\
		 e^{i\omega t} J_0 (t)^4
		},
		\qquad
	\{f_i\} = \pmat{  
		0\\
		0\\
		0\\
		0\\
		1
		},
		\nonumber
	\end{align}
	\begin{align}
	\{A_{ij}\}  &= \pmat{
		i\omega - \tfrac{4}{t}	& -4 & 0 & 0 & 0 \\
		1 & i\omega - \tfrac{3}{t}	& -3 & 0 & 0 \\
		0 & 2 & i\omega - \tfrac{2}{t}	& -2 & 0 \\
		0 & 0 & 3 & i\omega - \tfrac{1}{t}	& -1 \\
		0 & 0 & 0 & 4 & i\omega 
	}.
	\end{align}
Using Eqs.~\eqref{FiAnsatz}, \eqref{FiPrimeAnsatz},	 \eqref{LargeLinearSystem}, and \eqref{IFromFi} with $m=11$ collocation points gives the integral from $t=10$ to $t=100$ with absolute error about $10^{-6}$ (relative error about $10^{-3}$).  The intermediate quantities are illustrated in Fig.~\ref{levin}.

It is interesting to note that in the Levin method, the properties of the integrand are encapsulated in the $A_{ij} (t)$, so that the integrand is \emph{never} evaluated at any $t$ (unlike in traditional quadrature methods).  The special functions $e^{i\omega t}$, $J_0(t)$, and $J_0'(t)$ are only evaluated at $t=10$ and at $t=100$.  Thus, the computational time is dictated not by the number of integrand evaluations, but by the solution of the $mn$ simultaneous equations for the Chebyshev coefficients.

The above-described Levin method has some drawbacks in the present context:

\begin{itemize}
\item 
Integrating from $t=0$ to $t=\infty$ requires some extensions due to the singularity in $A_{ij} (t\rightarrow 0)$ and the infinite upper limit.
\item 
Solution of the collocation equations may be less controlled (with respect to roundoff accumulation) than performing quadrature on the smoothly decaying integrand in Eq.~\eqref{g2tauintegrals}.
\item 
Since the true $F_i(t)$ is smooth, the simplest way to improve collocation accuracy is to increase the number $m$ of Chebyshev polynomials, but this is restricted by the $O(m^3n^3)$ time complexity for solving the collocation equations.  
In contrast, Clenshaw-Curtis quadrature for the smoothly decaying integrand in Eq.~\eqref{g2tauintegrals}, which essentially involves integrating the Chebyshev interpolant, can be taken to arbitrarily high Chebyshev degree $m$, since the complexity scales only as $O(m)$ (or $O(m \ln m)$ if the quadrature weights are computed on the fly).
\item 
When $\omega$ lies at a van Hove singularity, the Levin method experiences problems due to resonance between the $e^{i\omega t}$ and $J_0(t)$ factors: the integrand acquires a non-oscillatory component that needs special treatment.
\end{itemize}

We observe an interesting parallel.  The complex-plane method for $G_\0(\omega)$ involves a sum of $d+1$ terms, and for $G_{r_1\dots r_d}(\omega)$ the sum involves $2^d$ terms.  The Levin method for $G_\0(\omega)$ requires $n=d+1$ basis functions, and for $G_{r_1\dots r_d}(\omega)$ it requires $n=2^d$ basis functions.  However, the complex method involves quadrature, which takes time linear in $d+1$ (or $2^d$), whereas solution of the Levin collocation equations takes time cubic in $d+1$ (or $2^d$).  Thus for large $d$ the Levin method is more expensive.

Taking all the above considerations into account, we feel that the complex-plane method is more suitable for evaluating hypercubic lattice Green functions of the form Eq.~\eqref{BesselFormula}.  Nevertheless, one's choice may be influenced by the availability of complex Hankel functions and quadrature routines in one's preferred language or software platform.  We have not attempted an exhaustive study of whether other methods for oscillatory integration\cite{filon1928,chung2000,iserles2005,olver2006,huybrechs2006,iserles2011} are able to deal with Eq.~\eqref{BesselFormula}.

\section{Selected values of G} 
The following values of the Green function for isotropic cubic and hypercubic lattices, i.e., with $\{\Omega_k\}=(1,1,1)$ and $(1,1,1,1)$ respectively, were computed using the complex-plane method implemented in Mathematica (see auxiliary file \texttt{HLGFPackage.nb}):
	\begin{align}
 	G_{000} (3) &= 0.50546201972  \nonumber\\
 	G_{100} (3) &=	-0.17212868638 \nonumber\\
 	G_{110} (3) &= 0.11038286738  \nonumber\\
 	G_{111} (3) &= -0.08715670880  \nonumber\\
 	G_{200} (3) &= 0.08577862908  \nonumber\\
 	G_{000} (0) &= -0.89644078878 i \nonumber\\
 	G_{100} (0) &= 0.33333333333 \nonumber\\
 	G_{110} (0) &= 0.18578752146 i \nonumber\\
 	G_{111} (0) &= -0.27566444771 \nonumber\\
 	G_{200} (0) &= 0.15329070292 i ,
	\end{align}
	\begin{align}
 	G_{0000} (4) &= 0.309866780462 \nonumber\\
 	G_{1000} (4) &= -0.05986678046 \nonumber\\
 	G_{1100} (4) &= 0.02542940754 \nonumber\\
 	G_{1110} (4) &= -0.01546809528 \nonumber\\
 	G_{1111} (4) &= 0.01118185767 \nonumber\\
 	G_{2000} (4) &= 0.01649101798 \nonumber\\
 	G_{0000} (0) &= -0.90272857832 i \nonumber\\
 	G_{1000} (0) &= 0.25000000000 \nonumber\\
 	G_{1100} (0) &= 0.15098515279 i \nonumber\\
 	G_{1110} (0) &= -0.10132118364 \nonumber\\
 	G_{1111} (0) &= -0.20025275758 i \nonumber\\
 	G_{2000} (0) &= -0.00318233840 i \nonumber\\
 	G_{0000} (1) &= 0.3726972107993 - 0.6681496264378 i \nonumber\\
 	G_{0000} (2) &= 0.5680714850367 - 0.3573566432144 i \nonumber\\	
 	G_{0000} (3) &= 0.4358824699995 - 0.1063899831047 i .
	\end{align}
As expected, $\Im G_\rrr(\omega)=0$ when $\left|\omega\right| \geq d$ (the density of states is zero outside the band).  By symmetry, $\Re G_\0(0)=0$.  The Green function obeys the discrete Helmholtz equation, 
$\omega G_\rrr (\omega) + \half \sum_{k=1}^d \sum_{\sigma=\pm 1} G_{\rrr + \sigma\eee_k} (\omega) = \delta_\rrr$, where $\{\eee_k\}$ are unit vectors and $\delta_\rrr=1$ if $\rrr=\0$ and zero otherwise.  At the band bottom ($\omega=-d$) the Green function obeys the discrete Laplace equation, so $\Im G_\rrr(-d)=0$ and $\Re G_\rrr(-d)<0 ~\forall \rrr$.  The Green function at the top of the band alternates in sign between adjacent sites.

The accuracy of the above results ($11$--$13$ digits) was limited partly by the large errors in the Bessel-related functions supplied by Mathematica 8.0.1.0.  

\section{Conclusions} 
We have presented an efficient method [Eqs.~\eqref{gsplit},\eqref{g1},\eqref{g2tauintegrals}], based on complex analysis, for calculating Green functions of $d$-dimensional anisotropic hypercubic lattices at arbitrary lattice vectors.  We also discuss the merits of an alternative approach using the Levin collocation method, but we provide reasons to prefer the complex-variable method.

\mysection{Acknowledgments} 
YLL is grateful to Jean-Michel Caillol for helpful discussions.


\end{document}